\documentclass[aps,onecolumn,showpacs,preprintnumbers,amsmath,amssymb,nofootinbib]{revtex4} \usepackage{graphicx}
\usepackage{epsfig}
\usepackage{dcolumn}
\usepackage{bm}
\usepackage[english]{babel}
\usepackage{amsmath}
\usepackage{amsfonts}
\usepackage{amssymb}
\usepackage{delarray}
\numberwithin{equation}{section}
\newcommand{\g}{\; \hat{\mbox{\rm g}}}
\newcommand{\e}{\; \hat{\mbox{\rm e}}}
\begin{document}
\title{QUANTUM GRAVITY IN HEISENBERG REPRESENTATION AND SELF--CONSISTENT THEORY OF GRAVITONS IN MACROSCOPIC SPACETIME}

\author{Grigory Vereshkov} \email{gveresh@gmail.com} \affiliation{Research Institute of Physics, Southern Federal University, Rostov-on-Don 344090, Russia \protect\\
Institute for Nuclear Research of the Russian Academy of Sciences‡}

\author{Leonid Marochnik} \email{lmarochn@umd.edu}
\affiliation{Physics Department, University of Maryland, College Park, MD 20742, USA}

\begin{abstract}

The first mathematically consistent exact equations of quantum gravity in the Heisenberg representation and Hamilton gauge are obtained. It is shown that the path integral over the canonical variables in the Hamilton gauge is mathematically equivalent to the operator equations of quantum theory of gravity with canonical rules of quantization of the gravitational and ghost fields. In its operator formulation, the theory can be used to calculate the graviton $S-$matrix as well as to describe the quantum evolution of macroscopic  system of gravitons in the non--stationary Universe or in the vicinity of relativistic objects. In the $S-$matrix case, the standard results are obtained. For problems of the second type, the original Heisenberg equations of quantum gravity are converted to a self--consistent system of equations for the metric of the macroscopic spacetime and Heisenberg operators of quantum fields. It is shown that conditions of the compatibility and internal consistency of this system of equations are performed without restrictions on the amplitude and wavelength of gravitons and ghosts.  The status of ghost fields in the various formulations of quantum theory of gravity is discussed.

\end{abstract}


\maketitle

\section{Introduction}\label{Int}

In the works \cite{1, 2} it is shown that the observed accelerated expansion of the Universe (the Dark Energy effect) \cite{3, 4} may be a consequence of {\it macroscopic effect of the quantum theory of gravity} which is the condensation of gravitons on the horizon scale of the non--stationary Universe. The theory of this macroscopic quantum effect is based only on the most general properties of quantum gravitational field in the non--stationary Universe, which are Bose--Einstein statistics, zero rest mass of gravitons, their conformal non-invariance and one--loop finiteness of quantum gravity off the graviton mass shell. (One--loop finiteness on the graviton mass shell was proved by 't Hooft and Veltman in 1974 \cite{5}.) In \cite{1}, three exact solutions of equations one--loop quantum gravity are presented. They are presented in terms of moments of spectral function of gravitons which is renormalized by ghosts. These solutions describe three different phase states of the condensate. One of these solutions describing the self--polarized condensate in the de Sitter space is represented in terms of field operators and state vectors. The effect of condensation is clearly demonstrated by the fact that the state vectors have the form of quantum coherent superposition of vectors corresponding to different occupation numbers of gravitons and ghosts having the same wavelength of the order of the radius of 4--curvature. Detailed analysis of all solutions is given in \cite{2}.

This work aims to show that the equations of one--loop quantum gravity obtained in \cite{1} are the only mathematically consistent of all in the available literature. As is known, the quantum gravity is non--renormalizable theory \cite{6}. At the same time, once again we emphasize that the pure one--loop quantum gravity (without matter fields) is finite. The macroscopic quantum effects of interest are already occurring in the one--loop approximation. Our program is to obtain the equations of the one--loop theory of the formally (in mathematical sense) exact equations of quantum gravity. We believe that only such a way to obtain the one--loop equations allows conducting the correct one--loop calculations that predict the condensation of gravitons on the horizon scale of the Universe.

For curvatures that are less than Planck curvature, quantum cosmology is represented as a theory of gravitons in the macroscopic spacetime with the self--consistent geometry. The quantum state of gravitons is determined by their interaction with a macroscopic field, and the macroscopic (background) geometry, in turn, depends on the state of gravitons. The background metric and graviton operator appearing in the self--consistent theory are extracted from the unified gravitational field, which initially satisfies exact equations of quantum gravity. The classical component of the unified field is a function of coordinates and time by definition. The quantum component of the same unified field is described by a tensor operator function, which also depends on coordinates and time. Under such a formulation of the problem, the original exact equations should be the operator equations of quantum theory of gravity in the Heisenberg  representation. The present work is devoted to obtaining such equations and their identity transformation to the self--consistent system of classical and quantum equations with no restrictions on the wavelength and amplitude of the graviton field.

The Heisenberg representation of quantum theory can be introduced only by its Hamilton formulation. For the theory with constraints, we deal with a generalized Hamilton formalism. The generalized Hamilton formalism of the theory of gravity was constructed by Dirac \cite{7,7a} and Arnowitt, Deser and Misner \cite{8}. There are also other formulations of theory that differ from \cite{7,7a,8} by parameterization of gravitational variables. We use the formalism proposed by Faddeev \cite{9} (see section 5 in the review \cite{10} or a book \cite{11}, \S \ 24). In the non--degenerate field theories and in the theories that allow to remove the degeneracy completely, the transition from the classical Hamilton theory to the quantum theory in the Heisenberg representation is reduced to a simple replacement of the classical Poisson brackets with the quantum commutators. Conducting of the same operations in the theory of gravity would be possible if the four local gauge conditions together with four equations of constraints in the theory, would completely determine physical degrees of freedom that are to be quantized. In the theory of gravity local gauges completely removing the degeneracy are unknown. Therefore, the question arises, {\it whether there is a Heisenberg representation of quantum gravity in local gauges, if these gauges are unable to separate the true gravitational field from the fields of inertia?} This paper gives a positive answer to this question. The important fact is that the transition from classical to quantum theory is not limited to the replacement of $C-$numeric functions with the operator functions satisfying the canonical commutation relations. Such a replacement must be accompanied by the appearance of the operator functions of Faddeev--Popov ghosts (see Section \ref{Hei}).

At the present time, an operator version of equations proposed in \cite{7,7a,8} is usually considered as the equations of quantum gravity in the Heisenberg representation. One should, however, bear in mind that local gauge conditions used in \cite{7,7a,8} do not remove the degeneracy completely, i.e. quantum fields of inertia are present in the equations of theory. Therefore, perturbative $S-$matrixes calculated in these theories do not satisfy the unitarity condition. As is known, the existence of this problem drew the attention of Feynman \cite{12} in the analysis of the theory in the local covariant gauge which is convenient for perturbative calculations. Faddeev and Popov \cite{13} suggested a regular method of unitary $S-$matrix calculation, based on introduction of the auxiliary ghost fields in the path integral. Any correct formulation of quantum theory of gravity should allow to calculate the graviton $S-$matrix as one of its objectives, and the result of this calculation should coincide with the Faddeev--Popov $S-$matrix.

Despite this fact, the operator Einstein equations in the Heisenberg representation with the local gauges (without ghosts) that do not remove the degeneracy completely and do not satisfy the test for unitarity $S-$matrix, were used in discussing of cosmological applications of quantum gravity, i.e. in the quantum  theory of gravitons in the non--stationary Universe (see, e.g., \cite{14, 15}).

In connection with the situation that exists in the scientific literature, we emphasize the following circumstance. Except the test on a unitarity of $S-$matrix, there is no other criterion of correctness of the equations of quantum gravity, and, as a consequence, the correctness of the results obtained from these equations. Therefore, the Hamilton equations in the Heisenberg representation with the canonical quantization rules must be strictly deduced from the exact equations of quantum gravity satisfying the above test. In this paper, we show that these equations do exist under condition that the Faddeev--Popov ghosts are among the elements of the Hamilton formalism of quantum gravity.

We obtain the Hamilton equations in the Heisenberg representation by identity transformation of the path integral over the canonical variables. Such an integral was proposed by Faddeev \cite{9} on the basis of the general theory of Hamilton systems with explicitly unsolvable constraints \cite{16}. The ghost sector is originally contained in this integral, and for this reason it will inevitably be present in the mathematically equivalent operator Hamilton equations.

The procedure for obtaining of these equations is as follows. In the first step, the gauge of Faddeev's path integral \cite{9} is specified such that the ghost sector corresponding  to it  permits the introduction of canonical variables of the ghost fields and the representation of ghost Lagrangian in the Hamilton form. In the second step, the standard definition of the operator of evolution is used, and the a transition is made from the gauged path integral to the canonical Hamilton formalism in the Heisenberg representation. The canonical operator quantization acquires the status of a method which is a mathematical equivalent to the functional integration over the extended phase space of gravitons and ghosts.

The Heisenberg representation in the quantum theory of gravity (outside of perturbation theory) exists in the synchronous gauge in the normal coordinates. This gauge is an analogue of Hamilton gauge in the Yang--Mills theory. Basically, the operator Hamilton formalism with the canonical quantization of gravitational and ghost fields is to solve problems in quantum cosmology but can be also used for the calculation of $S-$matrix where it leads to standard results.

\section{Classical Einstein equations in Hamilton gauge}

\subsection{Einstein equations in normal coordinates}

Upon obtaining of the exact equations of the theory of gravity  in the Hamilton form, the gravitational field can be regarded as the deviation of the metric from the metric of Minkowski space $\bar g_{ik} = {\text{diag}} (1, -1, -1, -1)$. Normal coordinates of gravitational field of $\hat\Psi_i^k$ are given by exponential parameterization of the density of the contravariant metric \cite{17}
\begin{equation}
\begin{array}{c}
\displaystyle \sqrt{-\hat{g}}\hat{g}^{ik} = \sqrt{-\bar g}\bar g^{il}\g_l^k\ ,
\\[5mm] \displaystyle \g_l^k\equiv (\exp{\hat\Psi})_l^k=  \delta_l^k + \hat\Psi_l^k + \frac{1}{2}\hat\Psi_l^m\hat\Psi_m^k+\ldots \ . \end{array}
\label{2.1}
\end{equation}
The density of the gravitational Lagrangian as a function of normal coordinates reads
\begin{equation}
\displaystyle {\mathcal L}_{grav}=-\frac{1}{2\varkappa}\sqrt{-\hat g}\hat g^{ik}\hat R_{ik}=\frac{1}{8\varkappa}\g_k^l\left(\hat \Psi_n^{m,\ k}\hat \Psi_{m,\ l}^n-\frac12\hat\Psi^{,\ k}\hat\Psi_{,\ l} -2\hat\Psi^{k,\ m}_n\hat\Psi^n_{m,\ l}\right)+...\ , \label{2.2}
\end{equation}
where dots denote a full derivative which does not contribute to the equations of motion. A variation of the action of the gravity theory over the normal coordinates leads to the Einstein equation
\begin{equation}
\begin{array}{c}
\displaystyle \sqrt{-\hat g}\hat g^{kl}\hat R_{il}\equiv\frac12\left(\g^{ml}\hat\Psi^{k}_{i,\ m}-\g^{km}\hat\Psi^{l}_{i,\ m}-\g^{ml}\Psi^k_{m,\ i} -\frac12 \delta_i^k\g^{ml}\hat \Psi_{,\ m}\right)_{,\ l}-
\\[5mm]
\displaystyle -\frac14\g^{kl}\left(\hat\Psi^n_{m,\ i} \hat\Psi_{n,\ l}^m- \frac12\hat\Psi_{,\ i} \hat\Psi_{,l}-2\hat\Psi_{l,\ m}^n\hat \Psi_{n,\ i}^m\right)=0\ .
\end{array}
\label{2.3}
\end{equation}
The same equations can be rewritten in the form
\begin{equation}
\begin{array}{c}
\displaystyle \hat{\mathcal E}_i^k\equiv\sqrt{-\hat g}\hat g^{kl}\hat R_{il}-\frac12\delta_i^k\sqrt{-\hat g}\hat g^{ml}\hat R_{ml}=0\ . \end{array}
\label{2.4}
\end{equation}
In accordance with the general properties of Einstein equations, $\sqrt{-\hat g}\hat g^{\beta l}\hat R_{\alpha l}=0$ are equations of motion ($\alpha,\ \beta=1,2,3$), and $\hat {\mathcal E}^0_i=0$ are equations of constraints. Note also the following form in which the Bianci identity can be presented:
\begin{equation}
\displaystyle \frac{\partial{\hat{\mathcal E}}_i^k}{\partial x^k}+\frac{1}{2}\frac{\partial\hat\Psi_k^l}{\partial x^i}\left(\hat{\mathcal E}_l^k-\frac12\delta_l^k\hat{\mathcal E}\right)\equiv 0\ .
\label{2.5}
\end{equation}

\subsection{Hamilton gauge, conservation laws and Hamiltonian formalism}\label{CHF}

The Hamilton gauge is given by additional conditions
\begin{equation}
\displaystyle \sqrt{-\hat g}\hat g^{00}=1\ , \qquad  \sqrt{-\hat   g}\hat g^{0\alpha}=0\ .
\label{2.6}
\end{equation}
In normal coordinates, the conditions (\ref{2.6}) are reduced to $\hat\Psi_0^i=0$. The specific of the Hamilton gauge is that the Bianchi identity (\ref{2.5}) takes the differential form of conservation laws over the solutions of the equations of motions $\sqrt{-\hat g}\hat g^{\beta l}\hat R_{\alpha l}=0$. It  reads:
\begin{equation}
\displaystyle \frac{\partial\hat{\mathcal E}_i^k}{\partial x^k}=0\ .
\label{2.7}
\end{equation}
After the gauge is applied, the Lagrangian and the equations of motions take the form:
\begin{equation}
\displaystyle {{\mathcal L}'}_{grav}=\frac{1}{8\varkappa}\left(\partial_0\hat\Psi_\nu^\mu\cdot \partial_0\hat\Psi_\mu^\nu-\frac12\partial_0\hat\Psi\cdot \partial_0\hat\Psi\right)+\frac{1}{8\varkappa}\g_\rho^\sigma\left(\hat \Psi_\nu^{\mu,\ \rho}\hat \Psi_{\mu,\ \sigma}^\nu-\frac12\hat\Psi^{,\ \rho}\hat\Psi_{,\ \sigma} -2\hat\Psi^{\rho,\ \mu}_\nu\hat\Psi^\nu_{\mu,\ \sigma}\right)\ ,
\label{2.8}
\end{equation}
\begin{equation}
\begin{array}{c}
\displaystyle \partial_0\partial_0\left(\hat\Psi_\alpha^\beta- \frac12\delta_\alpha^\beta\hat\Psi\right)+\partial_\gamma\left(\g^{\gamma}_\mu\hat\Psi^{\beta,\ \mu}_{\alpha}-\g^{\beta}_\mu \hat\Psi^{\gamma,\ \mu}_{\alpha}-\g^{\gamma}_\mu\Psi^{\beta\mu}_{\; ,\ \alpha} -\frac12 \delta_\alpha^\beta\g^{\gamma}_\mu\hat \Psi^{,\ \mu}\right)-
\\[5mm]
\displaystyle -\frac12\g^{\beta}_\gamma\left(\hat\Psi^\nu_{\mu,\ \alpha} \hat\Psi_{\nu,}^{\mu,\ \gamma}- \frac12\hat\Psi_{,\ \alpha} \hat\Psi^{,\ \gamma}-2\hat\Psi_{\nu,\ \mu}^\gamma\hat \Psi_{\mu,\ \alpha}^\nu\right)=0\ .
\end{array}
\label{2.9}
\end{equation}
The equations (\ref{2.9}) are obtained by variation of the gauged action $S'=\int {{\mathcal L}'}_{grav}d^4x$ over the variables $\hat\Psi_\beta^\alpha$. After the gauge is applied, the Lagrangian (\ref{2.8}) does not contain variables whose variations would produce equations of constraints $\hat {\mathcal E}_i^0=0$. However, these constraints are restored in solutions of the equations of motion (\ref {2.9}) by imposition of conditions on the numerical values of integrals of motion. As follows from (\ref{2.7}), first integrals of the equations of motion (\ref{2.9}) are of the form
\begin{equation}
\displaystyle \int\hat{\mathcal E}_i^0d^3x=C_i\ ,
\label{2.10}
\end{equation}
where $C_i$ are formally numerical constants whose values are fixed by initial and boundary conditions. The general covariant nature of the equations of constraints imposes the obvious limitations over values of $C_i$: only those initial and boundary conditions are allowable for the metric in the Hamilton gauge in which constants $C_i $, appearing in (\ref{2.10}), are equal to zero.

Thus, the formalism of the theory, based on a gauged Lagrangian (\ref{2.8}), differs from the formalism of non--degenerate field theories only by limitations $C_i=0$ that are related to the initial and boundary conditions but not to the equations. This fact allows to construct the Hamilton formalism easily. Generalized momentums are of a standard definition over functional derivatives:
\begin{equation}
\displaystyle \hat\pi_\mu^\nu=\frac{\delta {L'}_{grav}}{\delta(\partial_0\hat\Psi_\nu^\mu)}= \frac{1}{4\varkappa}\partial_0\left(\hat\Psi_\mu^\nu-\frac12\delta_\mu^\nu\hat\Psi\right)\ ,
\label{2.11}
\end{equation}
where ${L'}_{grav}=\int{{\mathcal L}'}_{grav}d^3x$. The gauged action is reduced to the form
\begin{equation}
\displaystyle S'=\int\left[\hat\pi_\mu^\nu\partial_0\hat\Psi_\nu^\mu-{\mathcal H}_{grav}(\hat\pi_\mu^\nu,\ \hat\Psi_\mu^\nu)\right]d^4x\ ,
\label{2.12}
\end{equation}
where
\begin{equation}
\displaystyle {\mathcal H}_{grav}=2\varkappa\left(\hat\pi_\nu^\mu \hat\pi_\mu^\nu-\hat\pi\hat\pi\right)-\frac{1}{8\varkappa}\g_\rho^\sigma\left(\hat \Psi_\nu^{\mu,\ \rho}\hat \Psi_{\mu,\ \sigma}^\nu-\frac12\hat\Psi^{,\ \rho}\hat\Psi_{,\ \sigma} -2\hat\Psi^{\rho,\ \mu}_\nu\hat\Psi^\nu_{\mu,\ \sigma}\right)
\label{2.13}
\end{equation}
is the density of canonical Hamiltonian by which the full Hamiltonian is calculated ${H}_{grav}=\int{\mathcal H}_{grav}d^3x$.  The variation of action (\ref{2.12}) over the generalized coordinates and momentums (as over independent variables) leads to the Hamilton system:
\begin{equation}
\begin{array}{c}
\displaystyle \partial_0\hat\pi_\alpha^\beta=-\frac{\delta H_{grav}}{\delta\hat\Psi_\beta^\alpha}=-\frac{1}{4\varkappa}\left[ \partial_\gamma\left(\g^{\gamma}_\mu\hat\Psi^{\beta,\ \mu}_{\alpha}-\g^{\beta}_\mu \hat\Psi^{\gamma,\ \mu}_{\alpha}-\g^{\gamma}_\mu\Psi^{\beta\mu}_{\; ,\ \alpha} -\frac12 \delta_\alpha^\beta\g^{\gamma}_\mu\hat \Psi^{,\ \mu}\right)\right.-
\\[5mm]
\displaystyle \left.-\frac12\g^{\beta}_\gamma\left(\hat\Psi^\nu_{\mu,\ \alpha} \hat\Psi_{\nu,}^{\mu,\ \gamma}- \frac12\hat\Psi_{,\ \alpha} \hat\Psi^{,\ \gamma}-2\hat\Psi_{\nu,\ \mu}^\gamma\hat \Psi_{\mu,\ \alpha}^\nu\right)\right]\ ,
\\[5mm] \displaystyle \partial_0\hat\Psi_\alpha^\beta=\frac{\delta H_{grav}}{\delta\hat\pi_\beta^\alpha}=4\varkappa\left(\pi_\alpha^\beta- \delta_\alpha^\beta\pi\right)\ .
\end{array}
\label{2.14}
\end{equation}
As it should be, the Hamilton formalism is mathematically equivalent to the Lagrange formalism, and this can be seen from the fact that the system of equations (\ref{2.14}) is easily reduced to the Lagrange equations (\ref{2.9}). The Hamilton constraint (\ref{2.10}) with $i = 0, \ C_0 = 0$ becomes the formula for calculating of the energy of the gravitational field through surface integral: \begin{equation}
\displaystyle H_{grav}\equiv\int{\mathcal H}_{grav}d^3x=\frac{1}{4\varkappa}\int\left(\g^\nu_\mu\hat\Psi_\nu^{\sigma,\ \mu}+ \g_\mu^\sigma\hat\Psi_\nu^{\mu,\ \nu}\right)dS_\sigma\ .
\label{2.15}
\end{equation}

\subsection{Inertia fields in Hamilton gauge}\label{IF}

The general formulae for infinitesimal transformations reads
\begin{equation}
\displaystyle \delta\sqrt{-\hat g}\hat g^{ik}=-\partial_l(\sqrt{-\hat g}\hat g^{ik}\eta^l)+\sqrt{-\hat g}\hat  g^{il}\partial_l\eta^k+ \sqrt{-\hat g}\hat g^{kl}\partial_l\eta^i\ .
\label{2.16}
\end{equation}
For the Hamilton gauge (\ref{2.6}), the expression (\ref{2.16}) leads to the equations for the parameters of residual transformations: \begin{equation}
\displaystyle \delta\sqrt{-\hat g}\hat g^{00}=-\partial_\alpha\eta^\alpha+ \partial_0\eta^0=0,\qquad \delta\sqrt{-\hat g}\hat g^{0\alpha}=\sqrt{-\hat g}\hat g^{\alpha\beta}\partial_\beta\eta^0+ \partial_0\eta^\alpha=0\ .
\label{2.17}
\end{equation}
The solution of equations (\ref{2.17}) is presented in the form
\begin{equation}
\begin{array}{c}
\displaystyle \eta^i_{iner}(t,{\bf x})=\eta_{scale}^i(t,{\bf x})+\eta_{rot}^i(t,{\bf x})+\eta_{wave}^i(t,{\bf x})\ ,
\end{array}
\label{2.18}
\end{equation}
where
\begin{equation}
\begin{array}{c}
\displaystyle \eta_{scale}^i(t,{\bf x})=\text{const}\cdot x^i\ ,\qquad \eta_{rot}^0=0\ ,\quad \eta_{rot}^\alpha({\bf x})=e^{\alpha\beta\gamma}\partial_\beta f_\gamma({\bf x})\ ,
\end{array}
\label{2.19}
\end{equation}
$f_\gamma({\bf x})$ is a function of spatial coordinates defined by arbitrarily;
\begin{equation}
\begin{array}{c}
\displaystyle \eta_{wave}^\alpha(t,{\bf x})=\bar g^{\alpha\beta}\partial_\beta\chi(t,{\bf x})\ , \qquad \chi(t,{\bf x})=\frac{1}{4\pi}\int\frac{\partial_0\eta_{wave}^0(t,{\bf x})}{|{\bf x}-{\bf x'}|}d^3x'\ ,
\end{array}
\label{2.20}
\end{equation}
$\eta_{wave}^0(t,{\bf x})$ is the solution of the d'Alambert wave equation
\begin{equation}
\displaystyle \partial_0\partial_0\eta_{wave}^0+ \partial_\alpha\sqrt{-\hat g}\hat  g^{\alpha\beta}\partial_\beta\eta_{wave}^0= 0\ . \label{2.21}
\end{equation}

Function (\ref{2.18}) defines the fields of inertia:
\begin{equation}
\begin{array}{c} \displaystyle \delta\left(\sqrt{-\hat g}\hat g^{\alpha\beta}\right)_{iner}=-\partial_l(\sqrt{-\hat g}\hat g^{\alpha\beta}\eta^l_{iner})+\sqrt{-\hat g}\hat  g^{\alpha\gamma}\partial_\gamma\eta^\beta_{iner}+ \sqrt{-\hat g}\hat g^{\beta\gamma}\partial_\gamma\eta^\alpha_{iner}=
\\[5mm]
\displaystyle =\delta\left(\sqrt{-\hat g}\hat g^{\alpha\beta}\right)_{scale}+\delta\left(\sqrt{-\hat g}\hat g^{\alpha\beta}\right)_{rot}+\delta\left(\sqrt{-\hat g}\hat g^{\alpha\beta}\right)_{wave}\ .
\end{array}
\label{2.22}
\end{equation}
According to (\ref{2.19}), fields of inertia $\delta\left(\sqrt{-\hat g}\hat g^{\alpha\beta}\right)_{scale}+\delta\left(\sqrt{-\hat g}\hat g^{\alpha\beta}\right)_{rot}$ are trivial in the sense that $\eta_{scale}^i(t, {\bf x})$ and $\eta_{rot}^i(t, {\bf x})$ do not interact with gravity. These fields describe the given motions of the reference frame. The motion corresponding to a field $\eta_{scale}^i(t, {\bf x})$ is a given stretch of coordinate grid, and the motion corresponding to a field $\eta_{rot}^i(t, {\bf x})$ is a given rotation. The lack of connection with the metric implies that these fields of inertia do not affect the solution of Einstein's equations; for this reason they can be excluded from consideration.

The function $\eta_{wave}^0(t, {\bf x})$ (satisfying the wave equation (\ref{2.21})) defines a nontrivial field of inertia $\delta\left(\sqrt{-\hat g}\hat g^{\alpha\beta}\right)_{wave}$. The nature of this field is following. The condition $\hat g_{0\alpha} = 0 $ synchronizes the clocks, and component $ \hat g_{00}$ sets the rate of a clock. According to (\ref{2.6}), the Hamilton gauge is synchronous, and the rate of a clock is rigidly connected with the dynamics of 3--volume: $\displaystyle \sqrt{\hat g_{00}}=\sqrt{\hat\gamma}$, where $\hat\gamma$ is the determinant of spatial metric. Effects of continuous clock adjustment to a new physical situation are reflected in the measured values of 3--dimensional component of the metric in the form of wave field of inertia. It is propagating with the same speed with which a clock located at the point of measurement is synchronized with the clocks located at the neighboring points.

Note that in the theory of gravity in the Hamilton gauge the non--trivial field of inertia is of one only independent internal degree of freedom.

\section{Path integral in canonical variables and operator equations in Heisenberg representation}\label{Hei}

The procedure of hamiltonization of classical theory of gravity in the Hamilton gauge is carried out in section \ref{CHF}. Its content and sequence of procedures exactly match to the hamiltonization procedure of Yang--Mills theory \cite{18}. Just as in \cite{18}, ch.III, \S \ 2, the Hamilton gauge first is introduced in the Lagrangian and the Lagrange equations of motion, then the theory is presented in the Hamilton form.

Differences between the Yang--Mills theory and the theory of gravity appear during the phase of quantization. According to \cite{18}, the fact that in the Yang--Mills theory the Hamilton gauge completely removes the degeneracy, lets go straight to the quantum Yang--Mills theory in the Heisenberg representation by replacing the classical Poisson brackets with quantum commutators. A transition to a path integral over the canonical variables is of the status of computing the Heisenberg operator of evolution. The completely analogous transition to the quantum theory of gravity is not possible because fields of inertia remain in the Hamilton gauge (see \ref{IF}). To compensate for contributions of interactions of inertia and gravitational fields to the observables, the ghost fields are needed, which occur only in the formalism of path integration. Taking into account this circumstance, the primary postulate (allowing the Heisenberg representation of quantum gravity) is introduced at the path integral level for the canonical variables in the Hamilton gauge:
\begin{equation}
\begin{array}{c}
\displaystyle \langle \mbox{\rm out}|\mbox{\rm in}\rangle= \int \exp\left\{i\int\left[\hat\pi_\mu^\nu\partial_0\hat\Psi_\nu^\mu-{\mathcal H}_{grav}(\hat\pi_\mu^\nu,\ \hat\Psi_\mu^\nu)\right]d^4x\right\}\left(\mbox{\rm det}\, \hat M^i_{\; k}\right)\prod_x\prod_{\mu\leqslant\nu}d\hat\pi_\mu^\nu d\hat\Psi_\mu^\nu\ ,
\end{array}
\label{3.1}
\end{equation} where
\begin{equation}
\hat M^i_{\; k}= \begin{pmatrix} \partial_0 &\qquad \qquad & -\partial_\alpha \\ \sqrt{-\hat g}\hat  g^{\alpha\beta}\partial_\beta && \delta_\beta^\alpha\partial_0 \end{pmatrix}
\label{3.2}
\end{equation}
is the operator of the system of equations (\ref{2.17}) for the parameters of residual transformations. The expression (\ref{3.1}) is obtained from Faddeev's path integral (see formula (5.6) in \cite{10} or formula (24.22) in \cite{11}). It is obtained after setting the Hamilton gauge, integration over $ \sqrt{- \hat g}\hat g^{0i}$ and introduction of normal coordinates and corresponding momentums. Note that a very simple structure of the gauged action in (\ref{3.1}) is due to the fact that in the classical theory in the Hamilton gauge the gauged variables are of the status of Lagrange multipliers to the classical equations of constraints.

The computation of the determinant of operator (\ref{3.2}) leads to
\begin{equation}
\displaystyle \mbox{\rm det}\, \hat M^i_{\; k}= \left[\mbox{\rm det}\; (\partial_0\partial_0+\partial_\alpha\sqrt{\hat g}\hat g^{\alpha\beta}\partial_\beta)\right]\times \left[\mbox{\rm det}\; \partial_0\right]\times \left[\mbox{\rm det}\; \partial_0\right]. \label{3.3}
\end{equation}
The structure (\ref{3.3}) completely corresponds to the classification of the fields of inertia in the Hamilton gauge. The localization $\mbox{\rm det}\, \hat M^i_{\; k}$ leads to the Lagrangian containing only one non--trivial ghost field, which does interact with the gravity. The direct localization leads to the path integral over the configuration space of ghost fields:
\begin{equation}
\displaystyle \mbox{\rm det}\, \hat M^i_{\; k}=\int\exp\left(i\int \mathcal{L}_{ghost}d^4x\right)\prod_xd\bar\theta d\theta\ , \label{3.4}
\end{equation}
where
\begin{equation}
\displaystyle \mathcal{L}_{ghost}=-\frac{1}{8\varkappa}\left(\partial_0\bar\theta\cdot\partial_0\theta+ \g^\nu_\mu\bar\theta^{,\ \mu}\theta_{,\ \nu}\right)
\label{3.5}
\end{equation}
is the density of the ghost Lagrangian. A path integral over the phase space is a mathematical equivalent to (\ref{3.4}). It reads: \begin{equation}
\displaystyle \mbox{\rm det}\, \hat M^i_{\; k}=\int\exp\left\{i\int\left[{\mathcal P}\cdot\partial_0\theta+ \partial_0\bar\theta\cdot{\bar{\mathcal P}}- {\mathcal H}_{ghost}({\mathcal P},\theta,{\bar{\mathcal P}},\bar\theta,\hat\Psi_\mu^\nu)\right]d^4x\right\} \prod_xd{\bar\theta} d{\bar{\mathcal P}}d\theta d{\mathcal P}
\label{3.6}
\end{equation}
where
\begin{equation}
\displaystyle {\mathcal H}_{ghost}({\mathcal P},\theta,{\bar{\mathcal P}},\bar\theta,\hat\Psi_\mu^\nu)=-8\varkappa{\mathcal P}{\bar{\mathcal P}}+\frac{1}{8\varkappa}\g_\mu^\nu{\bar\theta}_{,\ \nu}\theta^{,\ \mu}
\label{3.7}
\end{equation}
is the density of the ghost Hamiltonian. Substitution of (\ref{3.6}) to (\ref{3.1}) gives a path integral over the extended phase space of gravitational and ghost variables:
\begin{equation}
\begin{array}{c}
\displaystyle \langle \mbox{\rm out}|\mbox{\rm in}\rangle= \\[5mm] \displaystyle =\int \exp\left\{i\int\left[\hat\pi_\mu^\nu\cdot\partial_0\hat\Psi_\nu^\mu +{\mathcal P}\cdot\partial_0\theta+ \partial_0\bar\theta\cdot{\bar{\mathcal P}}- {\mathcal H}(\hat\pi_\mu^\nu,\ \hat\Psi_\mu^\nu, {\mathcal P},\theta,{\bar{\mathcal P}},\bar\theta)\right]d^4x\right\}\prod_x\prod_{\mu\leqslant\nu}d\hat\pi_\mu^\nu d\hat\Psi_\mu^\nu\prod_xd{\bar\theta} d{\bar{\mathcal P}}d\theta d{\mathcal P}\ ,
\\[5mm] \displaystyle {\mathcal H}(\hat\pi_\mu^\nu,\ \hat\Psi_\mu^\nu, {\mathcal P},\theta,{\bar{\mathcal P}},\bar\theta)= {\mathcal H}_{grav}(\hat\pi_\mu^\nu,\ \hat\Psi_\mu^\nu)+ {\mathcal H}_{ghost}({\mathcal P},\theta,{\bar{\mathcal P}},\bar\theta, \hat\Psi_\mu^\nu)\ .
\end{array}
\label{3.8}
\end{equation}
Expression (\ref{3.8}) is the standard definition of the matrix element of the operator of evolution that defines the dynamics of state vectors in the Schr$\ddot{\text{o}}$dinger representation or, equivalently, the dynamics of operators in the Heisenberg representation. Therefore (\ref{3.8}) allows immediately writing down the terms of the canonical quantization and equations of quantum gravity in the Heisenberg representation.

The graviton sector is:
\begin{equation}
\displaystyle \left[\hat\pi_\mu^\nu(t,{\bf x}),\ \hat\Psi_\rho^\sigma(t,{\bf x}')\right]_-=-i\delta_\mu^\sigma\delta_\rho^\nu\delta({\bf x}-{\bf x}')
\label{3.9}
\end{equation}
\begin{equation}
\begin{array}{c}
\displaystyle \partial_0\hat\pi_\alpha^\beta=i\left[H,\  \hat\pi_\alpha^\beta\right]_-=-\frac{1}{4\varkappa}\left[ \partial_\gamma\left(\g^{\gamma}_\mu\hat\Psi^{\beta,\ \mu}_{\alpha}-\g^{\beta}_\mu \hat\Psi^{\gamma,\ \mu}_{\alpha}-\g^{\gamma}_\mu\Psi^{\beta\mu}_{\; ,\ \alpha} -\frac12 \delta_\alpha^\beta\g^{\gamma}_\mu\hat \Psi^{,\ \mu}\right)\right.-
\\[5mm] \displaystyle
\left.-\frac12\g^{\beta}_\gamma\left(\hat\Psi^\nu_{\mu,\ \alpha} \hat\Psi_{\nu,}^{\mu,\ \gamma}- \frac12\hat\Psi_{,\ \alpha} \hat\Psi^{,\ \gamma}-2\hat\Psi_{\nu,\ \mu}^\gamma\hat \Psi_{\mu,\ \alpha}^\nu\right)\right]-\frac{1}{8\varkappa}\g^\beta_\gamma\left(\bar\theta_{,\ \alpha}\theta^{,\ \gamma}+\bar\theta^{,\ \gamma}\theta_{,\ \alpha}\right)\ ,
\\[5mm] \displaystyle
\partial_0\hat\Psi_\alpha^\beta=i\left[H,\  \hat\Psi_\alpha^\beta\right]_-=4\varkappa\left(\pi_\alpha^\beta- \delta_\alpha^\beta\pi\right)\ .
\end{array}
\label{3.10}
\end{equation}

The ghost sector is:
\begin{equation}
\displaystyle \left[ \theta(t,{\bf x}),\  \mathcal{P}(t,{\bf x}')\right]_+= \left[{\bar\theta}(t,{\bf x}),\ \bar{\mathcal{P}}(t,{\bf x}')\right]_+=i\delta({\bf x}-{\bf x}')\ ,
\label{3.11}
\end{equation}
\begin{equation}
\begin{array}{c}
\displaystyle \partial_0\mathcal{P}=i\left[H,\  \mathcal{P}\right]_-=\frac{1}{8\varkappa}\partial_\mu\g^\mu_\nu{\bar\theta}^{,\ \nu}\ ,  \qquad \partial_0\theta= i\left[H,\  \theta\right]_- =-8\varkappa\bar{\mathcal{P}}\ ,
\\[5mm] \displaystyle \partial_0\bar{\mathcal{P}}=i\left[H,\  \bar{\mathcal{P}}\right]_-  =\frac{1}{8\varkappa}\partial_\mu\g^\mu_\nu\theta^{,\ \nu}\ ,\qquad  \partial_0\bar\theta= i\left[H,\  \bar\theta\right]_- =-8\varkappa{\mathcal{P}}\ .
\end{array}
\label{3.12}
\end{equation}
In (\ref{3.10}), (\ref{3.12})
\begin{equation}
\displaystyle H=\int({\mathcal H}_{grav}+{\mathcal H}_{ghost})d^3x
\label{3.13}
\end{equation}
is the full canonical Hamiltonian of gravitational and ghost fields. The canonical quantization (\ref{3.9}), (\ref{3.11}) and solution of operator equations (\ref{3.10}), (\ref{3.12}) are procedures that are of mathematically equivalence to the computation of path integral (\ref{3.8}).

It is easy to check that differential conservation laws are satisfied for the solutions of equations (\ref{3.10}), (\ref{3.12})
\begin{equation}
\displaystyle \frac{\partial}{\partial x^k}\left(\hat{\mathcal E}_i^k-\varkappa\sqrt{-\hat g}\hat g^{kl}\hat T_{li}^{(ghost)}\right)=0\ ,
\label{3.14}
\end{equation}
where $\hat{\mathcal E}_i^k$ is the density of Einstein tensor,
\begin{equation}
\displaystyle \sqrt{-\hat g}\hat g^{kl}\hat T_{li}^{(ghost)}=-\frac{1}{4\varkappa}\left[\sqrt{-\hat g}\hat g^{kl}\left(\partial_l\bar\theta\cdot\partial_i\theta+ \partial_i\bar\theta\cdot\partial_l\theta\right)- \delta_i^k\sqrt{-\hat g}\hat g^{lm}\partial_l\bar\theta\cdot\partial_m\theta\right]
\label{3.15}
\end{equation}
is the density of the energy--momentum tensor of the ghosts. Integral conservation laws are presented as conditions imposed on the Heisenberg state vector, i.e. on the vector of initial quantum state:
\begin{equation}
\displaystyle \left[\int\left(\hat{\mathcal E}_i^0-\varkappa\sqrt{-\hat g}\hat g^{0l}\hat T_{li}^{(ghost)}\right)d^3x\right]|\Psi\rangle =0\ .
\label{3.16}
\end{equation}
The Hamilton constraint acquires a meaning of the equation for eigenvalues and eigenvectors of the total Hamiltonian of graviton and ghost fields:
\begin{equation}
\displaystyle \left(H_{grav}+H_{ghost}\right)|\Psi\rangle=E|\Psi\rangle
\label{3.17}
\end{equation}
A comparison of quantum Einstein's equations (\ref{3.10}) in the Hamilton gauge in the Heisenberg representation with the classical equations (\ref{2.14}) shows that the transition from classical to quantum theory is not limited to the replacement of $C-$numeric functions with the operator functions satisfying the canonical commutation relations. In the operator equations of quantum gravity, the operator functions of Faddeev--Popov ghosts automatically appear. They satisfy the canonical anti--commutation relations. The role of ghosts in the Heisenberg equations of motion (\ref{3.10}) is exactly the same as in the path integral (\ref{3.8}): the ghosts compensate for the effects of interaction of a true gravitational field with a field of inertia. However, in the complete system of equations (\ref{3.9}) --- (\ref{3.12}) the ghosts formally are of the status of additional dynamic variables, i.e. their status is equal to the status of dynamical variable of gravitational field.

At the operator level, the structure of equations of the theory in the Heisenberg representation with canonical quantization of gravitons and ghosts ensures the existence and carrying out routine operations on the transition from (\ref{3.9}) --- (\ref{3.12}) to the representation of interaction, and the construction of $S-$matrix in the framework of the theory of perturbations. In this theory, gravitons are described by the deviation of the metric from the metric of Minkowski space. Asymptotic states for gravitons are constructed over the vectors of the Fock space for the quanta of 3D--transverse tensor modes, and asymptotic states for the ghosts are set as vacuum states. Ghosts are present only in the Green function that compensates for the contribution of the wave field of inertia in the Green function of gravitons. The above $S-$matrix is identically coincides with $S-$matrix constructed from the path integral over canonical variables (\ref{3.8}) or from the path integral over the Lagrange variables (\ref{4.1}). Match of $S-$matrixes constructed in various ways is a direct illustration of the existence of the Heisenberg representation for the quantum theory of gravity in the Hamilton gauge. Application of this theory to the description of a macroscopic number of gravitons in the Riemannian space with a self--consistent geometry is discussed in section \ref{STG}.

\section{Path integral and Heisenberg representation in Lagrange variables}\label{Lag}

The quantum theory of gravity in the Heisenberg representation formulated in (\ref{3.9}) --- (\ref{3.12}) is mathematically equivalent to the calculation of gauged path integral over the Lagrange variables of gravitational and ghost fields:
\begin{equation}
\begin{array}{c}
\displaystyle  \langle \mbox{\rm out}|\mbox{\rm in}\rangle=
\\[5mm]\displaystyle
=\int \exp\left[-\frac{i}{2\varkappa}\int \sqrt{-\hat g}\hat g^{ik}\left(   \hat R_{ik}+\frac12\bar\theta_{,\ i}\theta_{,\ k}\right)d^4x\right]  \prod_x\left(\prod_i\delta(h_k\sqrt{-\hat g}\hat  g^{ik}-n^i)\right)(-\hat g)^{5/2}\prod_{i\leqslant k}d\hat g^{ik} d\bar\theta d\theta\ ,
\end{array}
\label{4.1}
\end{equation}
where $h_k=n^i=(1,0,0,0)$ are unit timelike vectors defining a Hamilton gauge. Of course, (\ref{4.1}) is a specific case of the Faddeev--Popov--DeWitt gauged path integral \cite{13,19,19a}. Structurally, (\ref{4.1}) is obtained from the gauge invariant expression \begin{equation}
\displaystyle \langle \mbox{\rm out}|\mbox{\rm in}\rangle= \frac{1}{\Omega}\int \exp\left[-\frac{i}{2\varkappa}\int \sqrt{-\hat g}\hat g^{ik}   \hat R_{ik}d^4x\right]  \prod_x(-\hat g)^{5/2}\prod_{i\leqslant k}d\hat g^{ik}
\label{4.2}
\end{equation}
through the expansion of unit in the Hamilton gauge and factorization of volume of the gauge group $\Omega$. In the context of existence of the Hamilton formalism, we note that (\ref{4.1}) can be regarded as the result of identity transformations of the expression (\ref{3.8}). We do not describe these transformations due to their rather obvious.

The expression (\ref{4.1}) clearly reflects the unique property of the Hamilton gauge: the effective Lagrangian of gauged theory is formally defined by the covariant expression
\begin{equation}
\displaystyle {\mathcal{L}}_{eff}=-\frac{1}{2\varkappa}\sqrt{-\hat g}\hat g^{ik}\left(   \hat R_{ik}+\frac12\bar\theta_{,\ i}\theta_{,\ k}\right)\ .
\label{4.3}
\end{equation}
Because of this property, the variation of action $S_{eff}=\int{\mathcal{L}}_{eff}d^4x$ over normal coordinates $\hat\Psi_k^i$ and Lagrange variables $\bar\theta,\ \theta$, leads to a system of generally covariant equations for the gravitational and Grassmann scalar fields:
\begin{equation}
\displaystyle \sqrt{-\hat g}\hat g^{kl}\hat R_{il}-\frac12\delta_i^k\sqrt{-\hat g}\hat g^{ml}\hat R_{ml}=-\frac{1}{4}\left[\sqrt{-\hat g}\hat g^{kl}\left(\partial_l\bar\theta\cdot\partial_i\theta+ \partial_i\bar\theta\cdot\partial_l\theta\right)- \delta_i^k\sqrt{-\hat g}\hat g^{lm}\partial_l\bar\theta\cdot\partial_m\theta\right]
\label{4.4}
\end{equation}
\begin{equation}
\displaystyle \partial_i\sqrt{-\hat g}\hat g^{ik}\partial_k\theta=0\ ,\qquad \partial_i\sqrt{-\hat g}\hat g^{ik}\partial_k\bar\theta=0\ .
\label{4.5}
\end{equation} The left--hand--side of Einstein's equations (\ref{4.4}) satisfies the Bianchi identities, and the energy--momentum tensor of Grassmann fields appearing in the right--hand--side is conservative over the equations of motion (\ref{4.5}). The existence of these differential identities allows separating Einstein's equations to the equations of motion of gravitational field and equations of constraints. It is also clear that after an explicit account of the Hamilton gauge, the gauged equations of motion are reduced to the form (\ref{3.10}), (\ref{3.12}). Thus, the Hamilton gauge ensures the existence of Heisenberg representation for the equations of quantum theory of gravity both in Hamilton and in the Lagrange form.

\section{Self--consistent theory of gravitons in macroscopic Riemannian space}\label{STG}

The exact equations in the Heisenberg representation in the Hamilton gauge (\ref{3.10}), (\ref{3.12}) together with postulates of canonical quantization of (\ref{3.9}), (\ref{3.11}) claim the status of the theory formulated only on the basis of the first principles of quantum gravity. We turn to the approximate theory of macroscopic system of gravitons in the macroscopic spacetime with the self--consistent geometry. In this theory, the interaction of gravitons with the classical gravitational field is taken into account exactly, and the perturbation theory over the amplitude of quantum fluctuations is only used when describing the graviton--graviton interaction.

The averaging of normal coordinates over the Heisenberg state vector in the general case provides non--zero $C-$numeric function: $\langle\Psi|\hat\Psi_i^k|\Psi\rangle=\Phi_i^k\ne 0$. A quantum fluctuation is defined as the difference between the operator and its average value: $\hat \psi_i^k = \hat \Psi_i^k-\Phi_i^k$. Substituting $\hat \Psi_i^k = \Phi_i^k + \hat \psi_i^k$ in the formula (\ref{2.1}) for the operator of the density of covariant metric gives
\begin{equation}
\begin{array}{c}
\displaystyle \sqrt{-\hat{g}}\hat{g}^{ik} =  \sqrt{-\bar g}\bar g^{il}[\exp{(\Phi+\hat\psi})]_l^k=  \sqrt{-g}g^{il}\e_l^k\ , \\[5mm]  \displaystyle \sqrt{-g}g^{il}=\sqrt{-\bar g}\bar g^{im}(\exp{\Phi})_m^k\ ,
\\[5mm]  \displaystyle
\e_l^k\equiv (\exp{\hat\psi})_l^k=\delta_l^k + \hat\psi_l^k + \frac{1}{2}\hat\psi_l^m\hat\psi_m^k+\ldots \ .
\end{array}
\label{5.1}
\end{equation}
A matrix exponential built over $C-$numeric functions $\Phi_i^k$, will be called the density of contravariant metric of the macroscopic spacetime. In this space, we normally enter the covariant metric $g_{ik}$, the connection $\Gamma_{ik}^l$, the curvature $R_{ik}$ and define covariant derivatives. The quantum fluctuations $\hat \psi_i^k$ are endowed by the properties of a tensor in the macroscopic spacetime. The Hamilton gauge $\hat \Psi_i^0 = 0$ is partitioned into gauge of the macroscopic metrics and gauge of the quantum field: \begin{equation}
\begin{array}{c}
\displaystyle \sqrt{-g}g^{00} =1\ ,\qquad \sqrt{-g}g^{0\alpha}=0\ ,
\\[5mm]  \displaystyle
\hat\psi_0^0=0\ ,\qquad \hat\psi_\alpha^0=0\ .
\end{array}
\label{5.2}
\end{equation}
Transformations of equations can be made in the Hamilton formalism (\ref{3.10}), (\ref{3.12}) as well as in the Lagrange formalism (\ref{4.4}), (\ref{4.5}). To elucidate the general properties of the theory, the Lagrange formalism is more convenient. Substituting (\ref{5.1}) in (\ref{4.4}) gives:
\begin{equation}
\begin{array}{c}
\displaystyle R_i^k-\frac12\delta_i^kR +\frac12\left(\hat\psi^{k\; ;l}_{i;l}-\hat\psi^{k\; ;l}_{l;i} -\hat\psi^{l;k}_{i\; ;l}+\delta_i^k\hat\psi_{m;l}^{l;m}\right)+\hat\psi^k_lR_i^l-\frac12\delta_i^k\hat\psi^m_lR_m^l=  \varkappa\hat T_i^k\ ,
\\[5mm] \displaystyle
\hat T_i^k=\hat T_{i(grav)}^k+\hat T_{i(ghost)}^k\ ,
\end{array}
\label{5.3}
\end{equation}
where
\begin{equation}
\begin{array}{c} \displaystyle \varkappa\hat T_{i(grav)}^k=\frac14\e^{kl}\left(\hat\psi^n_{m;\ i} \hat\psi_{n;\ l}^m- \frac12\hat\psi_{;\ i} \hat\psi_{;\ l}-2\hat\psi_{l;\ m}^n\hat \psi_{n;\ i}^m\right)-\frac18\delta_i^k\e^{rl}\left(\hat\psi^n_{m;\ r} \hat\psi_{n;\ l}^m- \frac12\hat\psi_{;\ r} \hat\psi_{;\ l}-2\hat\psi_{l;\ m}^n\hat \psi_{n;\ r}^m\right)-
\\[5mm] \displaystyle
-\frac12\left[\e_{(1)}^{lm}\left(\hat\psi^k_{i;\ m}- \hat\psi^k_{m;\ i}\right) - \e_{(1)}^{km} \hat\psi^l_{i;\ m}+\frac12\delta_i^k\left( \e_{(1)}^{mn} \hat\psi_{n;\ m}^l+ \e_{(1)}^{lm} \hat\psi_{m;\ n}^n\right)\right]_{;\ l}-\e_{(2)}^{kl}{R_{li} +\frac12\delta_i^k\e_{(2)}^{lm}}R_{ml}\ ,
\\[5mm] \displaystyle \e_{(1)}^{ik}=\e^{ik}- g^{ik}=\hat\psi^{ik}+\frac12\hat\psi^{il}\hat\psi_l^k+...\ , \qquad \displaystyle \e_{(2)}^{ik}=\e^{ik}- g^{ik}-\hat\psi^{ik}=\frac12\hat\psi^{il}\hat\psi_l^k+...
\end{array}
\label{5.4}
\end{equation}
is the energy--momentum tensor of gravitons;
\begin{equation}
\begin{array}{c}
\displaystyle \varkappa\hat T_{i(ghost)}^k=-\frac{1}{4}\left[\e^{kl}\left(\bar \theta_{;l}\theta_{;i}+\bar \theta_{;i}\theta_{;l}\right)-\delta_i^k\e^{ml}\bar\theta_{;m}\theta_{;l}\right]
\end{array}
\label{5.5}
\end{equation} is the energy--momentum tensor of ghosts.
The Lagrange equations for the ghosts read
\begin{equation} \displaystyle (\e^{ik}\theta_{;k})_{;i}=0,\qquad (\e^{ik}\bar\theta_{;k})_{;i}=0
\label{5.6}
\end{equation}
Semicolons in (\ref{5.3}) --- (\ref{5.6}) denote covariant derivatives in the macroscopic spacetime.

The transformation of shift $\hat\Psi_i^k = \Phi_i^k + \hat \psi_i^k$ is of physical meaning if the true degree of freedom of the quantized gravitational field are contained in the operator function $\hat \psi_i^k$, and the macroscopic gravitational field is self--consistent and serves as a way to describe collective interactions in the system consisting of a large (macroscopic) number of particles. The ability to carry out specific calculations in the approximation of a self--consistent field appears when symmetry properties of the macroscopic spacetime allow the introducing of a system of basis state vectors, and on this base to construct a state vector of gravitons and ghosts of a general form.
The Einstein equations for the macroscopic geometry are obtained from (\ref{5.3}) by averaging over the state vector:
\begin{equation}
\displaystyle R_i^k-\frac12\delta_i^kR =  \varkappa\langle\Psi|\hat T_i^k|\Psi\rangle\ .
\label{5.7}
\end{equation}
The equations for gravitons in the Heisenberg representation are obtained by subtraction (\ref{5.7}) from the exact equations (\ref{5.3}):
\begin{equation}
\begin{array}{c}
\displaystyle \hat L_i^k\equiv\frac12\left(\hat\psi^{k\; ;l}_{i;l}-\hat\psi^{k\; ;l}_{l;i} -\hat\psi^{l;k}_{i\; ;l}+\delta_i^k\hat\psi_{m;l}^{l;m}\right)+\hat\psi^k_lR_i^l-\frac12\delta_i^k\hat\psi^m_lR_m^l-  \varkappa\left(\hat T_i^k-\langle\Psi|\hat T_i^k|\Psi\rangle\right)=0\  .
\end{array}
\label{5.8}
\end{equation}
In terms of the path integral formalism, the existence of equations (\ref{5.7}), (\ref{5.8}), (\ref{5.6}) is provided by factorization of the measure of integration and consistent integration, first over the quantum fields of gravitons and ghosts, and then over the classical field. It can be seen from following. After an additive transformation $\hat\Psi_i^k = \Phi_i^k + \hat \psi_i^k$ is used, the measure of integration in normal coordinates and Hamilton gauge  is represented as:
\begin{equation}
\begin{array}{c}
\displaystyle \prod_x\left(\prod_i\delta(h_k\sqrt{-\hat g}\hat  g^{ik}-n^i)\right)(-\hat g)^{5/2}\prod_{i\leqslant k}d\hat g^{ik} d\bar\theta  d\theta=\prod_x\prod_i\delta(h_k\hat\Psi^{k}_i)\prod_{i\leqslant k}d\hat  \Psi^k_id\bar\theta d\theta= \\[5mm] \displaystyle = \left[\prod_x\prod_i\delta(h_k\Phi^{k}_i)\prod_{i\leqslant k}d  \Phi^k_i\right]\times\left[\prod_x\prod_i\delta(h_k\hat\psi^{k}_k)\prod_{i\leqslant k}d\hat  \psi^k_id\bar\theta  d\theta\right]\ .  \end{array}
\label{5.9}
\end{equation}
Formally, the exact integration of quantum fields $ \bar \theta, \ \theta, \ \hat \psi_i^k $ corresponds to the solution of quantum equations (\ref{5.6}), (\ref{5.8}), and the approximate integration of the classical field $ \Phi_i^k $ corresponds to the solution of the classical Einstein equations (\ref{5.7}).

The system of classical and quantum equations satisfies three differential identities. The left--hand--side of (\ref{5.7}) by definition satisfies the contracted Bianchi identities in the macroscopic spacetime. In the right-hand-side, the condition for conservation of the expectation value of  energy--momentum tensor of ghosts $\langle \Psi|\hat T_{i (ghost)}^k|\Psi\rangle_{; \ k} = 0$ satisfies to the equations of motion for the ghosts (\ref{5.6}). The expectation value of energy--momentum tensor of gravitons is of the similar properties. The calculation of a covariant divergence yields:
\begin{equation}
\displaystyle \langle\Psi|\hat T_{i(grav)}^k|\Psi\rangle_{;k}=\frac12\langle\Psi|\hat\psi^l_{k;i}\left(\hat L_l^k-\frac12\delta_l^k\hat L_m^m\right)|\Psi\rangle\ .
\label{5.10}
\end{equation}
As seen from (\ref{5.10}), the condition of conservation $ \langle \Psi|\hat T_{i (grav)}^k|\Psi\rangle_{; k} = 0 $ is performed over the quantum equations of motion (\ref{5.8}).

In the self--consistent theory of gravitons, the non--perturbative effect of the formation of the geometry of macroscopic spacetime is described in semi--classical/semi--quantum level: in the Einstein equations (\ref{5.7}) the source of the classical field is the expectation value of energy--momentum tensor of quantum fields. In the quantum sector, the order of a perturbation theory is given by maximum degree $ "n" $ in which the operators of quantum fields of gravitons and ghosts are in the quantum equations of motion (\ref{5.8}), (\ref{5.6}). In the same order of the perturbation theory, terms of the order of $ "n +1" $ are accounted in the energy--momentum tensor. As seen from (\ref{5.10}), the condition of conservativeness of the graviton energy--momentum tensor takes place in each order of perturbation theory separately. This fact is the basis for the application of the perturbation theory.

\section{Status of ghost fields in the various formulations of quantum gravity}\label{ghost}

A specific feature of quantum gravity in the Heisenberg representation is the inevitable presence of ghost fields in the canonical formalism of the theory. Formally (in mathematical sense), the ghost sector is a consequence of the lack of gauges fully removing the degeneracy with respect to the residual transformations of the group of diffeomorphisms. Physically, this means that in the theory of gravity the fields of inertia are described together with the true gravitational field. In the classical theory of gravity, this does not lead to any problems with interpretation because the fields of inertia and gravitational fields are locally separable.  Therefore, the deterministic evolution of each of these fields always allows finding invariant characteristics of local effects. The specificity of the quantum gravity lies in the fact that because of the lack of gauges completely removing the degeneracy, the existing formalism leads to the quantization of fields of both types (i.e., the uncertainty principle is turned out to be applied not only to the gravitational field but also to the field of inertia).

Of course, in the quantum gravity the question about the status of ghost fields arises as well as the question about the rules of computing the contribution of ghosts to the observables. Answers to these questions are contained in the formalism of the theory. The general answer is that in the quantum gravity the ghosts and fields of inertia together form a physical effect.  Details of this procedure depend, however, from the specific of a physical problem.

Let's start with the situation, which is well known and understood. The Faddeev--Popov path integral \cite{9,13} is initially defined as a method of calculating of the transition amplitude between asymptotic states in the framework of perturbation theory. Here should be noted that at the very formulation of the problem, the fields of inertia are unobservable in the asymptotic states, which is formally provided by appropriate boundary conditions. The impossibility of observing of the inertia fields is automatically accompanied by the lack of asymptotic states with the nonzero ghost occupation numbers. Formally, the lack of ghosts is ensured by the stability of the ghost vacuum and by the possibility of conducting of the various "expansions of unit"\ during the transformations of the gauged path integral. However, the ghosts as well as the fields of inertia are present in the virtual states in the region of interaction, and their quantitative role in the formation of the observable amplitude of the graviton scattering is given by the relevant Feynman diagrams. In various gauges, different inertial and ghost fields are appear but the total contribution of these fields in the graviton sector of $S-$matrix is gauge invariant.

In quantum gravity in the Heisenberg representation there is a possibility of the task, which physically is fully equivalent to the problem of the $S-$matrix calculation. Equations (\ref{3.9}) --- (\ref{3.12}) can be used to describe the gravitational field whose potentials are only weakly perturb the metric of Minkowski space. In this situation, one can also enter the asymptotic state, and then, specifying the flux of gravitons in the initial asymptotic state to calculate the flux of gravitons in the final asymptotic state. The ghosts together with the inertia fields influence the formation of finite flux of gravitons, and this effect is taken into account in the process of solution of the operator equations of motion.

There are two specifics in the solution of the problem described above. First, the Heisenberg representation in the quantum gravity as well as in the Yang - Mills theory exists only in special gauges, so that the quantum fields of inertia and ghost sector are not arbitrary. Second, in the equations (\ref{3.9}) --- (\ref{3.12}), there actually are no mathematical indications of any special status of ghost fields in contrast to the original path integral where they are. Formally, in these equations, the quantum wave field of ghosts acts as a second dynamic subsystem. Therefore, the exclusion of the ghosts of the asymptotic states is an additional condition. The mathematical consistency of the selection rule of the ghost--free asymptotic states is provided by the stability of the ghost vacuum in the theory of perturbations on the background of the Minkowski space.

We now turn to the most non--trivial situation, which is the role of the ghosts in the quantum theory of the graviton macroscopic system. Let us consider the equations (\ref{5.7}), (\ref{5.8}), (\ref{5.6}) in the one--loop approximation. In the macroscopic Einstein equations, the energy--momentum tensors of the gravitons and ghosts are quadratic in quantum fields:
\begin{equation}
\begin{array}{c}
\displaystyle R_i^k-\frac12\delta_i^kR =  \varkappa\langle\Psi|\hat T_{i(grav)}^k+\hat T_{i(ghost)}^k|\Psi\rangle\  ,
\\[5mm] \displaystyle
\varkappa\hat T_{i(grav)}^k=\frac14\left(\hat\psi^n_{m;\ i} \hat\psi_{n}^{m;\ k}- \frac12\hat\psi_{;\ i} \hat\psi^{;\ k}-2\hat\psi^{k;\ m}_n\hat \psi_{m;\ i}^n\right)-\frac18\delta_i^k\left(\hat\psi^{n;\ l}_{m} \hat\psi_{n;\ l}^m- \frac12\hat\psi^{;\ l} \hat\psi_{;\ l}-2\hat\psi_{l;\ m}^n\hat \psi_{n}^{m;\ l}\right)-
\\[5mm] \displaystyle
-\frac12\left[\hat\psi^l_m\left(\hat\psi^{k;\ m}_i- \hat\psi^{mk}_{\; \; \; \; ; i}\right) - \hat\psi^k_m \hat\psi^{l;\ m}_i+\frac12\delta_i^k\left( \hat\psi^n_m \hat\psi_{n}^{l;\ m}+ \hat\psi^l_m \hat\psi^{m;\ n}_n\right)\right]_{;\ l}-\frac12\hat\psi^k_m\hat\psi^m_lR_{i}^l +\frac14\delta_i^k\hat\psi^l_n\hat\psi^n_mR_{l}^m\ ,
\\[5mm] \displaystyle \varkappa T_{i(ghost)}^k=-\frac{1}{4}\left(\bar \theta^{;k}\theta_{;i}+\bar \theta_{;i}\theta^{;k}-\delta_i^k\bar\theta^{;l}\theta_{;l}\right)\ .
\end{array}
\label{6.1}
\end{equation}
The quantum fields of gravitons and ghosts are described by the following linear equations:
\begin{equation}
\begin{array}{c}
\displaystyle \frac12\left(\hat\psi^{k\; ;l}_{i;l}-\hat\psi^{k\; ;l}_{l;i} -\hat\psi^{l;k}_{i\; ;l}+\delta_i^k\hat\psi_{m;l}^{l;m}\right)+\hat\psi^k_lR_i^l-\frac12\delta_i^k\hat\psi^m_lR_m^l=0\  ,
\end{array}
\label{6.2}
\end{equation}
\begin{equation} \displaystyle \theta^{; k}_{\; ;\ k}=0,\qquad \bar\theta^{; k}_{\; ;\ k}=0
\label{6.3}
\end{equation}
{\it As is seen from (\ref{6.1}) --- (\ref{6.3}), the one--loop quantum gravity is not confined to the theory of free fields with the spin $J=2$ in the curved spacetime. The objects of the theory are necessarily two fields which are tensor field of spin $J =2$ and the scalar Grassmann field with spin $J=0$. Moreover, as is seen from the solutions of the equations (\ref{6.1}) --- (\ref{6.3}), the gravitons and ghosts must have an equal status of dynamical subsystems of macroscopic quantum gravity system} \cite{1,2}. This somewhat unusual structure of the theory (in comparison with the theory of physical fields with spin $J=0, \ 1/2, \ 1$) needs for additional comments.

The first thing that should be stressed is the fact that the inclusion of ghosts in the list of "physical" fields is the result of regular mathematical transformations. These transformations are as follows: the Faddeev path integral over the canonical variables (\ref{3.1}) $\longrightarrow$ equations of quantum gravity in the Heisenberg representation (\ref{3.9}) --- (\ref{3.12}) $\longrightarrow$ the allocation from the normal coordinates of the gravitational field of their mean values and the identity transformation of the exact equations in the Heisenberg representation to the self--consistent system of classical and quantum equations (\ref{5.7}), (\ref{5.8}), (\ref{5.6}). There is no question about the existence of such transformations because they obviously are in a formal mathematical sense. The question is whether or not these transformations lead to a new physical content of quantum theory of gravity in the last stage of computations (they are initially absent in the gauged path integral)?

The answer to this question is "Yes". The physical content of the theory is really changed but the reason for this change is absolutely transparent. The source path integral is defined as the mathematical object on the background of Minkowski space with the asymptotic boundary conditions and under the assumption of stability of graviton (and ghost) vacuum. From a mathematical point of view, the transition from this integral to the theory of the macroscopic quantum gravity system (\ref{5.7}), (\ref{5.8}), (\ref{5.6}) represents an extrapolation of the theory to a new physical area with the properties that differ from the area in which the theory was formulated initially.

First of all, it should be noted that the formation of the metric of the classical curved spacetime with the self--consistent geometry is a {\it significantly non--perturbative effect} of collective interactions in the macroscopic system. Cosmological applications of quantum gravity make the abandonment of the Minkowski space (appearing in the theory of $S-$matrix) inevitable in favor of the real existing curved spacetime. Direct consequences of a realistic formulation of the problem of describing of a macroscopic system of massless and conformal noninvariant quantum fields are {\it disappearance of asymptotic states and instability of the graviton--ghost vacuum.} Any observer located inside such a system (i.e., in fact, inside the real Universe) is in the area of interaction. It is rather obvious that for such an observer the non--perturbative effects of vacuum instability are priority subjects of research.

Next, you need to bear in mind that in General Relativity the observer is represented by the fields of inertia which are inevitably quantized (in quantum gravity) in the absence of gauges completely removing the degeneracy. Therefore, in terms of standard quantum gravity, among of {\it locally observable physical quantities}, there necessarily are the observables formed by quantum fields of inertia and ghosts jointly. The instability of the vacuum in the spectral region where the wavelength of quantum fields is comparable to the radius of 4--curvature does not allow to fix the quantum state of these fields by zero occupation numbers. Exactly for this reason, the ghost fields are beginning to perform the role of the second quantum subsystem which is dynamically equal to the graviton subsystem.

Thus, on the one hand, the ghosts perform their standard function in the theory of macroscopic quantum gravity systems, i.e. the ghost contributions together with the contributions of the inertia fields form the observables. But, on the other hand, this function occurs in the non--standard conditions of the absence of asymptotic states and the instability of the vacuum.

This specific of the ghost sector is fully reflected in the equations of one--loop quantum gravity (\ref{6.1}) --- (\ref{6.3}). In these equations, the interaction of gravitons with each other is taken into account through a self--consistent field. According to the most general concepts of the quantum theory of gravity (existing on today), the one--loop effects of the interaction can not be properly described without ghosts. In the equations (\ref{6.1}) --- (\ref{6.3}), the inertia fields are accumulated in the macroscopic self--consistent field. Therefore, the ghost contribution to the formation of observables is described by the obvious and only a way possible --- through the influence of ghosts on the self--consistent field. This contribution is taken into account through the presence of an averaged energy--momentum tensor of the ghosts in the macroscopic Einstein equations (\ref{6.1}).

The exact solutions of the equations (\ref{6.1}) --- (\ref{6.3}) describing macroscopic effects of quantum gravity in the homogeneous and isotropic non-stationary Universe are obtained in \cite{1, 2}. A possible role of graviton, ghost and instanton condensates in the formation of observable Dark Energy is also discussed in \cite{1, 2}.

\section{Conclusion}\label{Con}

In this paper we show that the quantum theory of gravity can be formulated as operator equations in the Hamilton gauge in the Heisenberg representation. The postulate of quantization in this theory is given by canonical commutation and anti--commutation relations for the generalized coordinates and momentums of gravitational and ghost fields. The equations of theory (\ref{3.9}) --- (\ref{3.12}) (or the equations obtained from these equations (\ref{5.7}), (\ref{5.8}), (\ref{5.6})) are used to describe a macroscopic system of gravitons forming a macroscopic spacetime with the self--consistent geometry.

There is the reason served as one of the main motives in writing this paper. During the last 30 years (since 1977 to 2008) many papers were published on the quantum theory of gravitons taking into account their backreaction on the cosmological background (quantum backreaction theory). To the best of our knowledge, there are no mathematically consistent papers correctly taking into account the structure of the existing quantum theory of gravity. In particular, the papers \cite{14, 15, 20, 21, 22} published in Physical Review D are erroneous. The typical errors are following:

1) Use of linear parameterization of the metric fluctuations leading to the non--self--consistent system of classical and quantum equations. This is a problem facing the authors of all works on the theory of gravitons, but they tried not to direct attention to the discussion of this matter.  Existence of the problem is exhaustively documented in the work \cite{23}. In the same paper it was clearly stated that the use of linear parameterization leads to the need for manual adjustment of the energy--momentum tensor of gravitons. Such a correction, although it restores the conservative nature of the energy--momentum tensor in the background space, is outside of mathematically consistent formalism of the theory. In the theory proposed in this paper, the gravitational field undergoes exponential parameterization corresponding to the use of normal coordinates in the functional space. With this parameterization, the energy-momentum tensor of gravitons satisfies the conservative condition automatically --- see (\ref{5.10}). The need to use normal coordinates obviously follows from the condition of bringing the path integral to the standard definition of the evolution operator --- see (\ref{3.1}) and (\ref{3.8}). The same result can be obtained in a purely classical self-consistent theory of gravitational waves. In this case, the consistency of wave and background equations must be ensured by the variational principle. In the frame of this scheme, detailed calculations are given in \cite{2} (Section II.D, formula (II.36), (II.37)).

2) Incorrect gauging (a gauge is imposed not on the full metric but on its fluctuations only). Such a method is a serious and obvious error if it is used in applications in which the macroscopic background has no a global 4--symmetry, permitting the separation of inertia fields. Formally, the mathematical nature of the error is as follows. The group of diffeomorphisms  acting in the original Riemannian space with the metric (\ref{2.1}), cannot be factorized into  the group of diffeomorphisms of the background space and the gauge group of gravitational waves after the separation in this metric  of background and wave modes according to (\ref{5.1}). However, this error does not affect the erroneous results of \cite{14, 15, 20, 21, 22} because in the quantum theory the main consequence of gauging is the emergence of the ghost sector, which in these works is still ignored.

3) It is an unpardonable error of \cite{14, 15, 20, 21, 22} to completely disregard the most important property of pure quantum gravity (without matter fields) which is its one-loop finiteness. Meanwhile, the one--loop finiteness is a rigorous and well--known t'Hooft and Veltman result \cite{5}. We do not know the reasons why the one-loop finiteness is not discussed and simply ignored in \cite{14, 15, 20, 21, 22}. We can only note that in the self-consistent theory of gravitons, incorrect formulations, using the operation of regularization and renormalization of one--loop divergences, have no mathematical meaning. As shown in \cite{2} (section XII), after introduction of one-loop counter-terms, the initial equation for gravitons is modified, which generates one--loop divergences of the new mathematical structure, and so ad infinitum. In other words, the one--loop renormalization of theory of gravitons in those versions of the theory which are used in \cite{14, 15, 20, 21, 22} is simply not possible, and for this reason, these versions do not exist in the mathematical sense. Self-consistent theory of gravitons exists only in the form in which it has, according to the t'Hooft and Veltman theorem, the property of one-loop finiteness.

4) Incorrect use (or ignore) the ghost sector. The lack of a one--loop finiteness in \cite{14, 15, 20, 21, 22} versions of the theory, is mathematically uniquely associated with either ignoring ghost sector \cite{14, 15, 20, 22}, or with a mathematically incorrect work with this sector \cite{21}. Ignoring ghost sector cannot be justified because in the theory of gravity there are no gauges that completely remove the degeneracy of the metric with respect to non--trivial residual gauge transformations. Apparently, this unconditional and, in general, a trivial fact was just outside the field of attention of the authors of \cite{14, 15, 20, 22}. Some attempt to fix the defect of theory, that is, formally introduces the ghosts in a self--consistent theory of gravitons, made in \cite{21}. This attempt, however, contains three types of mathematical errors. First, the gauge was applied not to the full metric but just for its graviton part which is mathematically erroneous (sees point "2" above). Second, the gauge used in \cite{21} does not provide bringing path integral to the form of the operator of evolution, which makes the transition to the Heisenberg representation impossible. Third, the authors of \cite{21}, apparently, to justify ignoring the ghosts in previous works, completely groundless claim that ghosts do not contribute to the observables.  For "justification" of this statement, it was drawn the condition BRST invariance of physical states. This third error itself consists of two mistakes. Even in theory of the graviton $S$-matrix, ghosts participate in the formation of scattering amplitudes, with the status of observables. The condition of BRST invariance refers to the rules of selection of vectors of asymptotic states that are vacuum ones over the occupation numbers of ghosts. In the self--consistent theory of gravitons in the non-stationary universe, there are fundamentally no asymptotic states corresponding to the stable vacuum. Therefore the BRST symmetry provides only a covariance of states but not their invariance. As a consequence, the ghosts are directly involved in the formation of observables. In the self-consistent theory of gravitons, they are geometric characteristics of the background space.

All these works completely lose any meaning after renormalization of gravitational Lagrangian by quadratic counterterms off the mass shell. This procedure modifies the original definition of the graviton and makes the theory mathematically inconsistent. This is an unavoidable and unacceptable internal contradiction of one--loop quantum gravity (see Section XII of \cite{2}.) Because of the finiteness of the theory off the mass shell, the incorrect counterterms simply does not arise with the right algorithm of computations.
In this paper, we showed that the key to obtaining  the correct equations of the one--loop theory (\ref{6.1}) --- (\ref {6.3}) is their correct derivation from the exact equations of quantum gravity. Our point of view is that for the solution of various problems in the quantum gravity, the source equations must be the same. Therefore, because of the lack of gauges completely removing degeneracy, the ghosts must be present in any representations of quantum gravity, and they participate in the formation of observables starting from the one--loop approximation. In the $S-$matrix theory, the ghost diagrams participate in the formation of the amplitude of graviton scattering.  In the self--consistent theory of gravitons (in the Heisenberg representation) the one--loop interactions are taken into account in the approximation of the self--consistent field. As a result, the ghosts will inevitably take participate in the formation of a self-consistent field via their own energy--momentum tensor on the right--hand--side of the macroscopic Einstein equations self--consistent with the quantum equations of motion. This mathematically inevitable fact was completely ignored in the previously published works.
\\

{\bf Acknowledgment.} We would like to express our deep appreciation to Ludwig D. Faddeev of the Steklov Mathematical Institute for graciously agreeing to read our manuscript and verifying the correctness of our approach before submission to the journal.
We are deeply grateful to Mikhail Shifman and Arkady Vainshtein of the University of Minnesota for discussions of the structure and content of the theory.  Also, we would like to express our deep appreciation to our friend and colleague Walter Sadowski for invaluable advise and help in the preparation of the manuscript.

\end{document}